\documentclass[12pt]{article}

\usepackage{amssymb}
\usepackage{amsmath}
\usepackage{graphicx}
\usepackage{color}

\def\be{\begin{align}}
\def\ee{\end{eqnarray}}
\def\ben{\begin{align*}}
\def\een{\end{eqnarray*}}

\def\orte{{\bf e}}

\def\eps{\varepsilon}

\def\dvec{{\bf d}}
\def\fvec{{\bf f}}
\def\jvec{{\bf j}}
\def\n{{\bf n}}
\def\rr{{\bf r}}
\def\v{{\bf v}}

\def\Avec{{\bf A}}
\def\Bvec{{\bf B}}
\def\Evec{{\bf E}}

\def\Svec{{\bf S}}
\def\Tvec{{\bf T}}

\def\calDvec{\mbox{\boldmath{$\cal{D}$}}}

\def\Sigmavec{\mbox{\boldmath{$\Sigma$}}}

\def\mdip{\mbox{\boldmath{$\mathfrak{m}$}}}
\def\calM{{\cal{M}}}

\def\intl{\int\limits}
\def\ointl{\oint\limits}

\def\divg{\mathop{\rm div}\nolimits}

\def\rot{\mathop{\rm rot}\nolimits}

\def\bnabla{\mbox{\boldmath{$\nabla$}}}

\def\pdd#1#2{\frac{\partial #1}{\partial #2}}

\begin{document}
\title{Radiation of the electromagnetic field beyond the dipole approximation}
\author{Andrij Rovenchak and Yuri Krynytskyi\\
Department for Theoretical Physics,\\
Ivan Franko National University of Lviv, Ukraine}

\date{July 9, 2018}

\maketitle

\abstract{
An expression for the intensity of the electromagnetic field radiation is derived up to the order next to the dipole approximation. Our approach is based on the fundamental equations from the introductory course of classical electrodynamics and the derivation is carried out using straightforward mathematical transformations. 

\textbf{Key words:} multipole expansion; radiation; higher multipole moments; anapole; toroidicity.
}

\section{Introduction}
Multipole expansion is a well-known technique applied in calculations of electromagnetic \cite{Frenkel:1926,French&Shimamoto:1953} and gravitational fields \cite{Thorne:1980,Blanchet:1998} at large distances from sources. In stationary cases, such as electrostatic problems, this approach is mainly linked to simple series expansions and certain symmetrization procedures \cite{Kocher:1978,Bezerra_etal:2012}. Treatment of magnetostatic problems requires more elaborate techniques beyond the lowest order \cite{Gray_etal:2010}. However, it becomes more tricky in the case of the electromagnetic field radiation. Moreover, in textbooks on classical electrodynamics one of the terms beyond the dipole approximation is typically omitted in expressions for the radiated power 
even if the detailed derivations are given (see, e.g., \cite{Jackson:1999,Griffiths:1999,Landau&Lifshitz:1971}; for an exception see \cite{Gray:1978}).

The goal of this article is to derive an expression for the radiated power in the approximation next to the dipole one using the basic equations of electrodynamics and the techniques from vector and tensor calculus. From the methodological point of view, our approach is advantageous comparing to the derivations based on gauge symmetries \cite{Dubovik&Tosunyan:1983} or solutions to the scattering problem \cite{Alaee_etal:2018}. Indeed, our derivation only requires knowledge of the fundamental relations from the introductory course of classical electrodynamics and involves straightforward mathematical transformations. In addition, the simplicity of our approach allows one to obtaining a correction to the dipole radiation sufficient for any practical purposes; more general derivations might be found in \cite{Gray_etal:2010,Dubovik&Tugushev:1990,Vrejoiu&Zus:2010,Radescu&Vaman:2012,Fernandez-Corbaton_etal:2017}.

The rest of the paper is organized as follows. An expression for the radiated power via the Poynting vector and magnetic field is obtained in Section~\ref{sec:Poynting}. The multipole expansion of the radiative part of the vector potential is given in Section~\ref{sec:A}. The detailed derivations of each contribution into the radiated power are presented in Section~\ref{sec:atg}. Brief discussion of the results is given in Section~\ref{sec:concl}.

\section{Poynting vector and radiated power}\label{sec:Poynting}
In a region of the space without charges and currents, the energy conservation law for the electromagnetic field in the differential form is
\be\label{eq:ECL}
\pdd{w}{t} + \divg \Svec = 0,
\end{align}
where $w$ is the energy density (we use the Gaussian units)
\be
w = \frac{E^2+B^2}{8\pi},
\end{align}
and $\Svec$ the energy flux density (the Poynting vector)
\be
\Svec = \frac{c}{4\pi} \Evec\times\Bvec.
\end{align}
 
Applying Gauss's theorem to Eq.~(\ref{eq:ECL}), the radiated power 
$$
I=-\frac{dW}{dt}\equiv\frac{d}{dt}\int w\,dV,
$$ 
can be reduced to the surface integral of the Poynting vector:
\be\label{I-multi}
I = \ointl_{\Sigma}\Svec\cdot d\Sigmavec =\intl_{\Omega=4\pi}|\Svec|r^2\,d\Omega, 
\end{align}
where on the r.h.s the integration is performed over the complete solid angle. From this expression, one can conclude that only those fields contribute to the radiation, which ensure the dependence $|\Svec|\propto 1/r^2$ as $r\to\infty$.

In the far zone, the magnetic and electric fields can be written as the sums
\be
\Evec = \Evec_0 +\Evec_1,\qquad
\Bvec = \Bvec_0 +\Bvec_1,
\end{align} 
where  the second terms ($\Evec_1$ and $\Bvec_1$) are proportional to $1/r$ and thus correspond to the radiation part of the field. Since the electric ($\Evec_1$) and magnetic ($\Bvec_1$) field vectors are orthogonal and equal by magnitude, it is sufficient to consider the norm of the Poynting vector for the radiation fields
\be
|\Svec_1|= \frac{c}{4\pi} \big|\Evec_1\times\Bvec_1\big|
= \frac{c}{4\pi} |\Bvec_1|^2.
\end{align}
Then the radiated power is given by
\be
I=\frac{c}{4\pi}\intl_{\Omega=4\pi}|\Bvec_1|^2r^2\,d\Omega.
\end{align}
The magnetic field is
\be
\Bvec(\rr,t) = \rot\Avec(\rr,t),
\end{align}
where $\Avec(\rr,t)$ is vector potential, which we calculate in the next Section.

\section{Vector potential}\label{sec:A}
\begin{figure}[h]
\centerline{\includegraphics[scale=1.2]{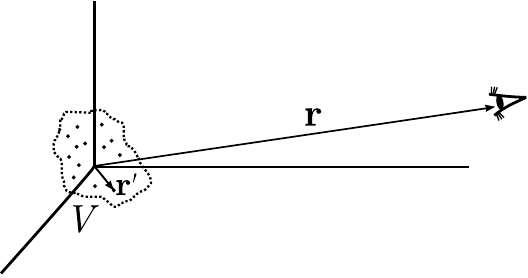}}
\caption{The system of charges is located near the origin and a distant observer sits at point $\rr$.}\label{fig:multipole}
\end{figure}

We consider a system of electric charges in a volume $V$ near the origin and assure that the observation point $\rr$ is far enough: $|\rr|\equiv r\gg V^{1/3}$, see Fig.~\ref{fig:multipole}. 

\bigskip

The vector potential of such a system in the Lorenz gauge is given by (cf. \cite[p.~408]{Jackson:1999}):
\be
\Avec(\rr,t) = {1\over c}\intl_VdV'\, \frac{1}{|\rr-\rr'|}\,
\jvec\left(\rr',t-{1\over c}|\rr-\rr'|\right),
\end{align}
where $dV'\equiv dx'\,dy'\,dz'$, and we have taken into account the retardation effects.

We write for $r'\ll r$ the following approximations over $r'/r$:
\be
\frac{1}{|\rr-\rr'|} &\simeq \frac{1}{r},\\
|\rr-\rr'| &\simeq r - \rr'\cdot\bnabla r = r - \frac{\rr'\cdot\rr}{r},
\end{align}
or, using the unit vector $\n = {\rr}/{r}$,
\be
|\rr-\rr'| \simeq r - \n\cdot\rr'.
\end{align}
Further terms bringing higher powers of $r$ in the denominators can be neglected when considering the radiation part of the field. 

Within this approximation, the vector potential yields
\be
\Avec(\rr,t) = {1\over cr}\intl_VdV'\,
\jvec\left(\rr',t-{r\over c}+{1\over c}\n\cdot\rr'\right).
\end{align}
It can be expanded in ${1\over c}\n\cdot\rr'$ using the Taylor series as follows:
\be\label{A_syst_Taylor}
\Avec(\rr,t) &= {1\over cr}\intl_VdV'\,
\jvec\left(\rr',t-{r\over c}\right)
+\frac{d}{dt} {1\over c^2r}\intl_VdV'\, (\n\cdot\rr')\,
\jvec\left(\rr',t-{r\over c}\right)+ \nonumber\\
&\qquad {}+ \frac{d^2}{dt^2} {1\over 2c^3r}\intl_VdV'\, (\n\cdot\rr')^2\,
\jvec\left(\rr',t-{r\over c}\right)+\ldots\;.
\end{align}

The current density for a system of point changes $e_i$ at $\rr_i$ moving with velocities $\v_i=\dot\rr_i$ can be written using the Dirac delta-function as follows:
\be
\jvec(\rr,\tau) = \sum_i e_i\v_i(\tau)\delta\big(\rr-\rr_i(\tau)\big),
\end{align}
where the retarded time $\tau = t-r/c$ was introduced for convenience.

The first term in expansion (\ref{A_syst_Taylor}) easily simplifies to the time derivative of the dipole moment:
\be
\intl_VdV'\, \jvec\left(\rr',t-{r\over c}\right) =
\sum_i e_i\v_i(\tau) = \frac{d}{d\tau}\sum_i e_i \rr_i(\tau) =
\frac{d}{d\tau} \dvec(\tau) = \dot\dvec(\tau).
\end{align}
Thus, we have in the leading order
\be
\Avec(\rr,t) = \frac{\dot\dvec\left(t-{r\over c}\right)}{cr} + \ldots\,.
\end{align}

To obtain the next-order correction, we make the following transformations:
\be\label{eq:A1}
&\intl_VdV'\, (\n\cdot\rr')\, \jvec\left(\rr',t-{r\over c}\right) =
\sum_i e_i\v_i(\tau) \intl_VdV'\, \n\cdot\rr'\, \delta\big(\rr'-\rr_i(\tau)\big) =
\nonumber \\
& = \sum_i e_i\v_i(\n\cdot\rr_i)  = \sum_i e_i\frac{d\rr_i}{d\tau}(\n\cdot\rr_i) =
\nonumber\\
& = {1\over2}\frac{d}{d\tau}\sum_i e_i\rr_i(\n\cdot\rr_i) +
{1\over2}\sum_i e_i\Big\{\v_i(\n\cdot\rr_i)-\rr_i(\n\cdot\v_i)\Big\} =
\nonumber \\
& = {1\over2}\frac{d}{d\tau}\sum_i e_i\rr_i(\n\cdot\rr_i) +
{1\over2}\sum_i e_i\, \n\times(\v_i\times\rr_i).
\end{align}
As we will see further, in the radiation parts of the field, the vector potential appears  only as cross product with the unit vector $\n$. Thus, one can add to $\Avec$ an arbitrary vector proportional to $\n$ without changing the results. It is convenient to add to the first term of (\ref{eq:A1}):
\ben
{1\over2}\frac{d}{d\tau}\sum_i e_i\rr_i(\n\cdot\rr_i)
 \quad\to\quad
{1\over2}\frac{d}{d\tau}\sum_i e_i\left\{\rr_i(\n\cdot\rr_i)
-\n\frac{r_i^2}{3}\right\}.
\end{align*}
We will further work with the vector potential $\Avec$ shifted as described above. The sum over $i$ is a contraction of the electric quadrupole moment tensor
\ben
Q_{jk} =
\sum_i e_i \left\{x_j^{(i)} x_k^{(i)} - \frac{r_i^2}3\delta_{jk}
\right\}
\end{align*}
with the unit vector $\n$, yielding some vector $\calDvec$ with components
\be
\mathcal{D}_j = \sum_k Q_{jk} n_k.
\end{align}
The remaining part of this correction involves the cross product of the magnetic dipole moment
\be
\mdip = \frac{1}{2c}\sum_i e_i\,\rr_i\times\v_i
\end{align}
with the unit vector $\n$.

The last term, which is written explicitly in (\ref{A_syst_Taylor}), can be transformed as follows:
\be
&\frac{d^2}{dt^2} {1\over 2c^3r}\intl_VdV'\, (\n\cdot\rr')^2\,
\jvec\left(\rr',t-{r\over c}\right) =\nonumber\\
&\qquad =
{1\over 2c^3r} \frac{d^2}{dt^2} \intl_VdV'\, \orte_i j_i\left(\rr',t-{r\over c}\right) n_kx'_k n_l x'_l =\nonumber\\
&\qquad = {1\over 2c^2r} \ddot\calM_{ikl}\left(\rr',t-{r\over c}\right) n_k n_l \orte_i,
\end{align}
where $\orte_i$ are the unit vectors of the Cartesian coordinate system and the summation over repeating indices is implied. The third-rank current quadrupole tensor is defined as:
\be
\calM_{ikl} = \frac1c \intl_VdV'\, j_i x'_k x'_l.
\end{align}
Note that we do not attempt to make this tensor traceless and just retain the main $x'_k x'_l$ term, which is sufficient for the purposes of further derivations.

Collecting all contributions in Eq.~(\ref{A_syst_Taylor}), we arrive at the vector potential in the following form:
\be\label{A-radiation-multi}
\Avec(\rr,t) = \frac{\dot\dvec}{cr} + \frac{\dot\mdip\times\n}{cr}
+\frac{1}{2c^2r}\ddot\calDvec + \frac{1}{2c^2r}\ddot\calM_{ikl} n_k n_l \orte_i,
\end{align}
where all the quantities are evaluated at the retarded time $\tau = t-r/c$. 
The obtained expression might be compared, e.g., to the radiation part of the vector potential in \cite[p.~17]{Raab&deLange:2005}.

\section{Bringing it all together}\label{sec:atg}
The next step is to calculate the magnetic field $\Bvec$ using expression (\ref{A-radiation-multi}) for the vector potential. We have:
\ben
\Bvec &= \rot\Avec = \bnabla\times\left[
\frac1{cr}\dot\dvec\left(t-\frac{r}{c}\right)+
\frac1{cr}\,\dot\mdip\left(t-\frac{r}{c}\right)\times\n+{}\right.\\
&\qquad\qquad\quad{}\left.+\frac1{2c^2r}\ddot\calDvec\left(t-\frac{r}{c}\right)
+ \frac{1}{2c^2r}\ddot\calM_{ikl}\left(t-\frac{r}{c}\right) n_k n_l \orte_i
\right].
\end{align*}
where we have written the time argument explicitly for convenience.

In order to retain the radiation part $\Bvec_1\propto 1/r$, it is sufficient to keep solely the terms where the nabla operator acts on the time argument only:
\ben
\bnabla\times \fvec\left(t-\frac{r}{c}\right) = 
\bnabla\left(t-\frac{r}{c}\right) \times \dot\fvec\left(t-\frac{r}{c}\right)=
\frac{1}{c}\,\dot\fvec\left(t-\frac{r}{c}\right)\times\n.
\end{align*}
This yields
\be\label{eq:B1}
\Bvec_1 = \frac{\ddot\dvec\times\n}{c^2r} +
\frac{(\ddot\mdip\times\n)\times\n}{c^2r} +
\frac{\dddot\calDvec\times\n}{2c^3r} + 
\frac{1}{2c^3r}\dddot\calM_{ikl}n_kn_l \,\eps_{irs}n_r\orte_s.
\end{align}

The four terms in Eq.~(\ref{eq:B1}) correspond to the electric dipole, magnetic dipole, electric quadrupole, and current quadrupole term, respectively. Explicitly this reads
\be
\Bvec_1 = 
\Bvec_{\rm d}+\Bvec_{\rm m}+\Bvec_{\rm Q}+\Bvec_{\cal M},
\end{align}
where (mind the $1/c$ in the definitions of the magnetic moments!)
\be
&\Bvec_{\rm d} = \frac{\ddot\dvec\times\n}{c^2r} \propto \frac{1}{c^2},\\[6pt]
&\Bvec_{\rm m} = \frac{\n\times(\n\times\ddot\mdip)}{c^2r} \propto \frac{1}{c^3},\\[6pt]
&\Bvec_{\rm Q} = \frac{\dddot\calDvec\times\n}{2c^3r} \propto \frac{1}{c^3},\\[6pt]
&\Bvec_{\cal M} = \frac{1}{2c^3r}\dddot\calM_{ikl} n_kn_l\,\eps_{irs}n_r\orte_s \propto \frac{1}{c^4},\label{BQm-def}
\end{align}

Thus, the magnitude of the Poynting vector contains the following terms:
\be\label{eq:S-full}
|\Svec_1|&=\frac{1}{4\pi}\Big(
\underbrace{c|\Bvec_{\rm d}|^2}_{\propto 1/c^3}+
\underbrace{2c\,\Bvec_{\rm d}\cdot\Bvec_{\rm m}}_{\propto 1/c^4}+
\underbrace{2c\,\Bvec_{\rm d}\cdot\Bvec_{\rm Q}}_{\propto 1/c^4}+\\[6pt]
&{}
+\underbrace{c|\Bvec_{\rm m}|^2}_{\propto 1/c^5}+
\underbrace{c|\Bvec_{\rm Q}|^2}_{\propto 1/c^5}+
\underbrace{2c\,\Bvec_{\rm m}\cdot\Bvec_{\rm Q}}_{\propto 1/c^5}+
\underbrace{2c\,\Bvec_{\rm d}\cdot\Bvec_{\cal M}}_{\propto 1/c^5}
\Big)+{\cal{O}}(c^{-6}).\nonumber
\end{align}

\subsection{Electric and magnetic dipole radiation}\label{sec:IdIm}
The leading contribution to the radiated power is given by the electric dipole term. It reads
\be
I_{\rm d} = \frac{c}{4\pi}\intl_{\Omega=4\pi} |\Bvec_{\rm d}|^2\,r^2\,d\Omega
=\frac{1}{4\pi c^3} \intl_{\Omega=4\pi} |\ddot\dvec\times\n|^2\,d\Omega.
\end{align}
Assuming that $\ddot\dvec$ is directed along the $Oz$ axis and that the $\theta$ angle of the spherical coordinate system is that between $\ddot\dvec$ and $\n$, we obtain
\be\label{eq:Id}
I_{\rm d} =\frac{\ddot\dvec^2}{4\pi c^3} \intl_{\Omega=4\pi} \sin^2\theta\,d\Omega = 
\frac{\ddot\dvec^2}{4\pi c^3} \intl_0^{2\pi}d\phi\intl_0^\pi d\theta\,\sin^3\theta = \frac{2\ddot\dvec^2}{3 c^3}\propto \frac{1}{c^3}.
\end{align}

In a similar fashion we can calculate the magnetic dipole contribution:
\be
I_{\rm m} = \frac{c}{4\pi}\intl_{\Omega=4\pi} |\Bvec_{\rm m}|^2\,r^2\,d\Omega
=\frac{1}{4\pi c^3} \intl_{\Omega=4\pi} |(\ddot\mdip\times\n)\times\n|^2\,d\Omega
\end{align}
considering $\theta$ as the angle between $\ddot\mdip$ and $\n$. We thus see that
\be\label{eq:Im}
I_{\rm m} = \frac{2\ddot\mdip^2}{3c^3} \propto \frac{1}{c^5}.
\end{align}

\subsection{Terms with zero contributions}

The product of the electric dipole and magnetic dipole terms $(\Bvec_{\rm d}\cdot\Bvec_{\rm m})$ contains the scalar product
\ben
(\ddot\dvec\times\n)\cdot[(\ddot\mdip\times\n)\times\n] &=
(\ddot\dvec\times\n)\cdot[\n(\n\cdot\ddot\mdip)-\ddot\mdip] =\\
&= (\ddot\dvec\cdot\underbrace{(\n\times\n)}_{=0})(\n\cdot\ddot\mdip) - (\ddot\dvec\times\n)\cdot\ddot\mdip =
\n\cdot(\ddot\dvec\times\ddot\mdip).\nonumber
\end{align*}
Integration of the respective term in Eq.~(\ref{I-multi}) using spherical coordinates with $\theta$ corresponding to the angle between $(\ddot\dvec\times\ddot\mdip)$ and $\n$ yields zero due to
\ben
\intl_0^\pi \sin\theta\cos\theta \,d\theta = 0.
\end{align*}

Similar considerations apply to the product of the electric dipole and the electric quadrupole terms $(\Bvec_{\rm d}\cdot\Bvec_{\rm Q})$. It contains $(\ddot\dvec\times\n)\cdot(\dddot\calDvec\times\n)$, which can be transformed as follows:
\ben
(\ddot\dvec\times\n)\cdot(\dddot\calDvec\times\n) =
\ddot\dvec\cdot[\n\times(\dddot\calDvec\times\n)] =
\ddot\dvec\cdot\dddot\calDvec - (\n\cdot\ddot\dvec)(\n\cdot\dddot\calDvec).
\end{align*}
Bearing in mind the definition of $\calDvec$, we obtain:
\ben
&\ddot\dvec\cdot\dddot\calDvec \equiv \ddot d_i \dddot{\mathcal{D}}_i =
\ddot d_i \dddot Q_{ij} n_j,\\
&(\n\cdot\ddot\dvec)(\n\cdot\dddot\calDvec) \equiv
\ddot d_i n_i \dddot{\mathcal{D}}_{j} n_j =
\ddot d_i \dddot Q_{jk}n_i n_j n_k ,
\end{align*}
where the summation over repeating indices is implied, as before.
It is easy to show that the integration of the $n_i$ components over the complete solid angle gives zero, e.g.:
\ben
\intl_{\Omega = 4\pi} n_x \,d\Omega =
\intl_0^{2\pi} d\psi \intl_0^\pi d\theta\,\sin\theta\cdot
\underbrace{\sin \theta\cos\psi}_{n_x=x/r} = 0.
\end{align*}
The same holds true for triple products:
\ben
\intl_{\Omega = 4\pi} n_i n_j n_k \,d\Omega = 0.
\end{align*}
This means that there is no contribution into the radiated power from the product of the electric dipole and quadrupole terms.

In fact, there is a common reason for the above two contributions being zero: they contain products of an odd number of $n_i$  components. Such expressions always yield zero upon integration over the complete solid angle. This is a consequence of independence of such an integral of the choice of axes orientation. The product of $n_i$ is an odd-rank symmetric tensor, and the only tensor of this type invariant under axes rotation is zero. The same rationale would apply, e.g., to the products of $(\Bvec_{\rm m}\cdot \Bvec_{\mathcal{M}})$ and $(\Bvec_{\rm Q}\cdot \Bvec_{\mathcal{M}})$, which appear in higher orders of expansion of the radiated power.

Consider now the product of the magnetic dipole and the electric quadrupole terms $(\Bvec_{\rm m}\cdot\Bvec_{\rm Q})$. The expression $(\dddot\calDvec\times\n)\cdot[(\ddot\mdip\times\n)\times\n]$ can be transformed as
\ben
(\dddot\calDvec\times\n)\cdot[(\ddot\mdip\times\n)\times\n] = \n\cdot(\dddot\calDvec\times\ddot\mdip).
\end{align*}
Since $\calDvec$ is a vector obtained as a contraction of the electric quadrupole moment tensor $Q_{ij}$ with the unit vector $\n = \rr/r$, we obtain:
\ben
\n\cdot(\dddot\calDvec\times\ddot\mdip) =
n_i \eps_{ijk} \dddot{\mathcal{D}}_j \mathfrak{m}_k =
 \eps_{ijk} \dddot{Q}_{jl} n_l n_i\mathfrak{m}_k.
\end{align*}
The integrals of two unit vectors over the solid angle can be shown to equal
\ben
\intl_{\Omega = 4\pi} n_l n_i \,d\Omega = \frac{4\pi}{3}\,\delta_{li}.
\end{align*}
The Kronecker delta contracts with $\dddot{Q}_{jl}$, hence the above term is proportional to
\ben
\eps_{ijk} \dddot{Q}_{jl} \delta_{li}\mathfrak{m}_k = \eps_{ijk} \dddot{Q}_{ji}\mathfrak{m}_k =0
\end{align*}
since this is the double contraction of the antisymmetric Levi-Civita symbol with symmetric quadrupole tensor derivatives. 

Such a double contraction of the Levi-Civita symbol with symmetric tensors will also yield zero contributions for terms of a similar nature in higher orders of expansion.

\subsection{Electric quadrupole radiation}

The square of the electric quadrupole term $|\dddot\calDvec\times\n|^2$ can be transformed as follows:
\ben
(\dddot\calDvec\times\n)\cdot(\dddot\calDvec\times\n) &=
\dddot\calDvec\cdot\dddot\calDvec - (\n\cdot\dddot\calDvec)(\n\cdot\dddot\calDvec)
=\dddot {\mathcal{D}}_i \dddot {\mathcal{D}}_i -
n_i \dddot {\mathcal{D}}_i n_j \dddot {\mathcal{D}}_j =\\
&= \dddot Q_{ij}n_j\dddot Q_{ik}n_k -
n_i \dddot Q_{ik}n_k n_j \dddot Q_{jl}n_l
\end{align*}
Thus, the power of the electric quadrupole radiation is equal to
\be\label{IQ1}
I_{\rm Q} = \frac{c}{4\pi}\frac1{4c^6}\intl_{\Omega=4\pi}
\left\{
\dddot Q_{ij}\dddot Q_{ik} n_jn_k -
 \dddot Q_{ik}\dddot Q_{jl} n_i n_j n_k n_l 
\right\} \,d\Omega.
\end{align}
The integral of four unit vectors is \cite[p.~415]{Jackson:1999}:
\be
\intl_{\Omega=4\pi} {n_i n_j n_k n_l}\,d\Omega =
\frac{4\pi}{15}
(\delta_{ij}\delta_{kl} + \delta_{ik}\delta_{jl} + \delta_{il}\delta_{jk}).
\end{align}
Taking into account that the electric quadrupole moment tensor is traceless, $Q_{ii}=0$,  Eq.~(\ref{IQ1}) reduces to
\be\label{IQ2}
I_{\rm Q} = \frac1{20c^5}\dddot Q_{ij}\dddot Q_{ij}.
\end{align}
The summation over indices $i,j$ is implied in this expression. 

It is worth noting that the definition of the quadruple moment tensor differ in the literature. For instance, using the definition
\ben
\widetilde Q_{jk} =
\sum_i e_i \left\{3x_j^{(i)} x_k^{(i)} - r_{(i)}^2\delta_{jk}
\right\}
\end{align*}
we obtain a different multiplier in the expression for the quadrupole radiation power, viz.
\be\label{IQ2'}
I_{\rm Q} = \frac1{180c^5}\dddot{\widetilde Q}_{ij} \dddot{\widetilde Q}_{ij}.
\end{align}

\subsection{Anapole radiation}
The product of the electric dipole term
\ben
\Bvec_{\rm d} = \frac{1}{c^2r}\,\ddot\dvec\times\n = \frac{1}{c^2r} \eps_{pqt}\ddot d_p n_q \orte_t,
\end{align*}
with the current quadrupole term (\ref{BQm-def}) yields
\ben
2\,\Bvec_{\rm d}\cdot\Bvec_{\cal M} &= 2\frac{1}{c^2r} \eps_{pqt} \ddot d_p n_q\orte_t \cdot \frac{1}{2c^3r}\dddot\calM_{ikl} n_kn_l\,\eps_{irs}n_r\orte_s = \\
&= \frac{1}{c^5r^2} \underbrace{\orte_t\cdot\orte_s}_{=\delta_{ts}} \eps_{pqt}\eps_{irs}n_kn_ln_qn_r \ddot d_p \dddot\calM_{ikl} = \\
&= \frac{1}{c^5r^2} \ddot d_p \dddot\calM_{ikl} \eps_{pqt}\eps_{irt}n_kn_ln_qn_r.
\end{align*}
It involves the contraction of the Levi-Civita symbols
\ben
\eps_{pqt}\eps_{irt} = \delta_{pi}\delta_{qr} - \delta_{pr}\delta_{qi}
\end{align*}
resulting in
\ben
2\,\Bvec_{\rm d}\cdot\Bvec_{\cal M} = \frac{1}{c^5r^2}
\left( \ddot d_i \dddot\calM_{ikl} n_k n_l - \ddot d_q \dddot\calM_{ikl} n_kn_ln_qn_i\right),
\end{align*}
where the square of a unit vector $n_rn_r=1$.

We will denote the contribution to the radiated power originating from this term as $I_{\rm A}$. It equals
\ben
I_{\rm A} &= \frac{c}{4\pi}
\frac{1}{c^5} \intl_{\Omega=4\pi}
\left( \ddot d_i \dddot\calM_{ikl} {n_k n_l \mathstrut} -
\ddot d_q \dddot\calM_{ikl} {n_kn_ln_qn_i \mathstrut}\right) d\Omega.
\end{align*}
Performing the transformations similar as in the previous subsection when dealing with the electric quadrupole radiation, we arrive at 
\ben
I_{\rm A}
&= \frac{1}{c^4}
\left(\frac{4}{15} \dddot\calM_{ikk} \ddot d_i -
\frac{2}{15} \dddot\calM_{iik} \ddot d_k \right)=
-\frac{2}{15c^4}\left(\dddot\calM_{kki} - 2 \dddot\calM_{ikk}\right)\ddot d_i.
\end{align*}
This result can be presented in a more convenient form introducing a vector
\be
\Tvec(\tau) &= \frac{1}{10} \left(\calM_{kki} - 2 \calM_{ikk}\right)\orte_i =
\nonumber\\[6pt]
&=
\frac{1}{10c} \intl_V dV'\,
\bigg\{
\big(\jvec(\rr',\tau)\cdot\rr'\big)\rr' - 2r'^2\, \jvec(\rr',\tau),
\bigg\}.
\end{align}
known as the \textit{anapole\footnote{The term was proposed by Zel'dovich following the suggestion by Kompaneets \cite{Zeldovich:1957}.} moment} or \textit{toroidicity}. The former name is due to the fact that this expression has no correspondence in the multipole expansion of static electric and magnetic fields. On the other hand, a toroidal solenoid produces a field, which can be described by $\Tvec$. Note that terms corresponding to the toroidal moments are sometimes included in the definitions of the dynamic electric multipole moments, and corresponding terms given for the radiated power \cite{Gray:1978,Fernandez-Corbaton_etal:2017,Alaee_etal:2018}.

The power of the anapole radiation is thus
\be\label{eq:IA}
I_{\rm A}=
- \frac{4}{3c^4}\,\dddot\Tvec\cdot\ddot\dvec.
\end{align}
In summary, the radiated power up to $1/c^5$ is given by:
\be
I &= I_{\rm d}+I_{\rm m}+I_{\rm Q}+I_{\rm A} \nonumber\\
&=
\frac{2\ddot\dvec^2}{3 c^3}
+ \frac{2\ddot\mdip^2}{3 c^3}
+ \frac1{20c^5}\dddot Q_{ij}\dddot Q_{ij}
- \frac{4}{3c^4}\,\dddot\Tvec\cdot\ddot\dvec.
\end{align}
Since the $1/c$ factor enters both the magnetic moment and toroidicity, the last three terms are proportional to $1/c^5$, being thus the $1/c^2$ lower correction to the leading term, i.e., the (electric) dipole approximation. Any subsequent terms in the expansion of the vector potential would yield corrections of the order $1/c^7$ and higher. From the considerations of symmetric properties of tensor products it can be shown that even powers of $1/c$ would be missing from the expansion, just like the $1/c^4$ terms in Eq.~(\ref{eq:S-full}).

\section{Discussion}\label{sec:concl}
It is not difficult to see why the anapole term is often neglected when considering the radiated power. The reason is that $I_{\rm A}$ contains the second time derivative of the electric dipole moment $\ddot\dvec$, which also defines the electric dipole radiated power $I_{\rm d}$. Usually, the latter is sufficient as the principal approximation and the calculation of the corrections is required only if $I_{\rm d}=0$. But in this case $\ddot\dvec=0$ and hence $I_{\rm A}=0$ as well. This means that the anapole term becomes relevant only if the radiated power requires a higher precision than the dipole term alone. 

In the static case, the electromagnetic field of the torus is zero outside the system but time-dependent distributions of charges and currents generate the electric field, in particular, with the radiation pattern of a dipole \cite{Carron:1995}. Such a moment is known as toroidicity, toroidal dipole or anapole moment. In higher approximations, other toroidal multipoles appear as well \cite{Gray_etal:2010,Dubovik&Tugushev:1990,Vrejoiu&Zus:2010,Fernandez-Corbaton_etal:2017}.

Applying the same strategy as in in derivations of the electric and magnetic dipole radiation in Sec.~\ref{sec:IdIm}, it is straightforward to show that the radiation of the torus is given by
\be
I_{\rm torus} =\frac{2\dddot\Tvec^2}{3c^5}\propto \frac{1}{c^7}.
\end{align}
This contribution, together with the electric dipole (\ref{eq:Id}) and anapole (\ref{eq:IA}) radiation, completes the square:
\be
I_{\rm d + t}=
\frac{2}{3c^3}\left(\ddot\dvec-\frac{1}{c}\dddot\Tvec\right)^2 =
\frac{2\ddot\dvec^2}{3c^3}-\frac{4}{3c^4}\,\dddot\Tvec\cdot\ddot\dvec+
\frac{2\dddot\Tvec^2}{3c^5},
\end{align}
cf. also \cite{Dubovik&Tugushev:1990}.

In summary, we have obtained an expression for the power of the electromagnetic field radiation in the approximation next to the dipole one, i.\,e., with terms proportional to $1/c^5$, which is the accuracy sufficient for comparisons with most present-day measurements.

\section*{Acknowledgment}
The authors are grateful to Volodymyr Tkachuk, Svyatoslav Kondrat, and Tim Brookes for critical reading of the manuscript and for the feedback. The authors appreciate comments from the anonymous Referees.


\end{document}